\documentclass[manuscript]{acmart}
\AtBeginDocument{%
  }

\setcopyright{acmlicensed}
\copyrightyear{2025}
\acmYear{2025}
\acmDOI{XXXXXXX.XXXXXXX}
\acmConference[Conference acronym 'XX]{Make sure to enter the correct
  conference title from your rights confirmation email}{June 03--05,
  2018}{Woodstock, NY}
\acmISBN{978-1-4503-XXXX-X/2018/06}

\usepackage{algorithm}
\usepackage{pifont}
\usepackage{booktabs}
\usepackage{tabularx}
\usepackage{multirow}
\usepackage{listings}
\usepackage{threeparttable}

\usepackage{amsmath}
\usepackage{amssymb}
\usepackage{algpseudocode}
\usepackage{cleveref}
\crefname{section}{§}{§§}
\usepackage{xcolor}
\usepackage{tcolorbox}
\definecolor{verylightgray}{rgb}{.97,.97,.97}
\usepackage{newfloat}
\lstdefinelanguage{Solidity}{
	keywords=[1]{anonymous, assembly, assert, balance, break, call, callcode, case, catch, class, constant, continue, constructor, contract, debugger, default, delegatecall, delete, do, else, emit, event, experimental, export, external, false, finally, for, function, gas, if, implements, import, in, indexed, instanceof, interface, internal, is, length, library, log0, log1, log2, log3, log4, memory, modifier, new, payable, pragma, private, protected, public, pure, push, require, return, returns, revert, selfdestruct, send, solidity, storage, struct, suicide, super, switch, then, this, throw, transfer, true, try, typeof, using, value, view, while, with, addmod, ecrecover, keccak256, mulmod, ripemd160, sha256, sha3}, 
	keywordstyle=[1]\color{blue}\bfseries,
	keywords=[2]{address, bool, byte, bytes, bytes1, bytes2, bytes3, bytes4, bytes5, bytes6, bytes7, bytes8, bytes9, bytes10, bytes11, bytes12, bytes13, bytes14, bytes15, bytes16, bytes17, bytes18, bytes19, bytes20, bytes21, bytes22, bytes23, bytes24, bytes25, bytes26, bytes27, bytes28, bytes29, bytes30, bytes31, bytes32, enum, int, int8, int16, int24, int32, int40, int48, int56, int64, int72, int80, int88, int96, int104, int112, int120, int128, int136, int144, int152, int160, int168, int176, int184, int192, int200, int208, int216, int224, int232, int240, int248, int256, mapping, string, uint, uint8, uint16, uint24, uint32, uint40, uint48, uint56, uint64, uint72, uint80, uint88, uint96, uint104, uint112, uint120, uint128, uint136, uint144, uint152, uint160, uint168, uint176, uint184, uint192, uint200, uint208, uint216, uint224, uint232, uint240, uint248, uint256, var, void, ether, finney, szabo, wei, days, hours, minutes, seconds, weeks, years},	
	keywordstyle=[2]\color{teal}\bfseries,
	keywords=[3]{block, blockhash, coinbase, difficulty, gaslimit, number, timestamp, msg, data, gas, sender, sig, value, now, tx, gasprice, origin},	
	keywordstyle=[3]\color{violet}\bfseries,
	identifierstyle=\color{black},
	sensitive=true,
	comment=[l]{//},
	morecomment=[s]{/*}{*/},
	commentstyle=\color{gray}\ttfamily,
	stringstyle=\color{red}\ttfamily,
	morestring=[b]',
	morestring=[b]"
}

\begin{document}

\title{No More Hidden Pitfalls? Exposing Smart Contract Bad Practices with LLM-Powered Hybrid Analysis}


\author{Xiaoqi Li}
\email{csxqli@ieee.org}
\affiliation{%
  \institution{Hainan University}
  \city{Haikou}
  \state{Hainan Province}
  \country{China}
}
\author{Zongwei Li}
\authornote{Corresponding author}
\email{lizw1017@hainanu.edu.cn}
\affiliation{%
  \institution{Hainan University}
  \city{Haikou}
  \state{Hainan Province}
  \country{China}
}
\author{Wenkai Li}
\email{cswkli@hainanu.edu.cn}
\affiliation{%
  \institution{Hainan University}
  \city{Haikou}
  \state{Hainan Province}
  \country{China}
}
\author{Yuqing Zhang}
\email{zhangyq@nipc.org.cn}
\affiliation{%
  \institution{University of Chinese Academy of Sciences}
  \city{Beijing}
  \postcode{100049}
  \country{China}
}

\author{Xin Wang}
\email{24210839000020@hainanu.edu.cn}

\affiliation{%
  \institution{Hainan University}
  \city{Haikou}
  \state{Hainan Province}
  \country{China}
}

\renewcommand{\shortauthors}{X. Li et al.}

\begin{abstract}
As the Ethereum platform continues to mature and gain widespread usage, it is crucial to maintain high standards of smart contract writing practices. While bad practices in smart contracts may not directly lead to security issues, they elevate the risk of encountering problems. Therefore, to understand and avoid these bad practices, this paper introduces the first systematic study of bad practices in smart contracts, delving into over 47 specific issues. Specifically, we propose \textsc{SCALM}, an LLM-powered framework featuring two methodological innovations: (1) A hybrid architecture that combines context-aware function-level slicing with knowledge-enhanced semantic reasoning via extensible vectorized pattern matching. (2) A multi-layer reasoning verification system connects low-level code patterns with high-level security principles through syntax, design patterns, and architecture analysis. Our extensive experiments using multiple LLMs and datasets have shown that \textsc{SCALM} outperforms existing tools in detecting bad practices in smart contracts.
\end{abstract}




\maketitle
\section{Introduction}
With the widespread use of blockchain technology, smart contracts have become an important part of the blockchain ecosystem \cite{Sharma_2023_review,huang2025comparative,caiEnablingCompleteAtomicity2024}. Smart contracts are computer programs that automatically execute contract terms, controlling assets and operations on the chain \cite{hanOSwapPreservingAtomicity2026}. However, due to their public and immutable code, smart contracts have become a significant target for attackers \cite{barboni2022smart}. A total of 464 security incidents occurred in 2023, resulting in losses of up to \$2.486 billion \cite{slowmist}. The most significant attack occurred on September 23rd when Mixin Network's cloud service provider database was attacked, involving approximately \$200 million.

\textbf{Bad practices} refer to poor coding habits or design decisions in the development of smart contracts \cite{reyes2023continuous,ding2025comprehensive,wu2025security}. We categorize bad practices into two types: (1) \textit{Security-related bad practices}, which include actual vulnerabilities that may lead to security incidents (e.g., reentrancy); and (2) \textit{Quality-related bad practices}, which encompass code quality issues that affect maintainability, efficiency, and design quality without directly resulting in exploitable vulnerabilities (e.g., code duplication). While security-related bad practices could potentially lead to future security threats, both types have significant impacts: they can cause performance problems, increase security risks, lead to unpredictable code behavior \cite{10.1145/3711901}, and create hidden economic dangers due to the disruption of regular smart contract activities \cite{sharma2023mixed}.

Currently, the security audit of smart contracts mainly relies on manual code review and automated tools \cite{10.1145/3641846,yang2025multi,wang2024ContractsentryStaticAnalysis,grossmanPracticalVerificationSmart2024}. However, these methods have their limitations \cite{10.1145/3674725,10.1145/3628160,li2025penetrating,ayubSoundAnalysisMigration2024}. Manual code review is inefficient and prone to overlook subtle security vulnerabilities. Existing automated tools primarily rely on pattern matching, which cannot accurately detect complex security issues. 
Moreover, the types of vulnerabilities that these tools can detect are usually relatively limited and may not be able to identify all potential security problems in smart contracts \cite{chaliasos2024smart}. 
To achieve a comprehensive audit, multiple tools may be required, each covering different aspects of security. 
Therefore, effectively detecting and preventing security issues in smart contracts remains an important issue that needs to be solved.

To address these challenges, we propose \textsc{SCALM} (\textbf{S}mart \textbf{C}ontract \textbf{A}udit \textbf{L}anguage \textbf{M}odel), a framework with two innovations for smart contract auditing \cite{chenDemystifyingInvariantEffectiveness2024}. \textbf{First}, a static analysis module constructs an extensible knowledge base by extracting code patterns, converting them into semantic vectors via an embedding model, and dynamically updating the repository with newly identified bad practices through automated SWC-ID annotation \cite{aguilarSmartContractFamilies2024,arceriSoundConstructionEVM2024}. \textbf{Second}, a multi-layer reasoning verification system combines Retrieval-Augmented Generation (RAG) with Step-Back prompting, enabling hierarchical reasoning from syntax checks to architectural risk analysis. This framework detects both explicit vulnerabilities and latent design flaws, generating structured audit reports. This approach generates detailed audit reports that document identified issues, assess risk scores, and provide concrete remediation suggestions, achieving improved detection accuracy compared to existing tools while maintaining practical applicability for developers.
Our contributions are as follows:
\begin{itemize}
\item To the best of our knowledge, we provide the first systematic study of bad practices in smart contracts and conduct an in-depth discussion and analysis on 35 security-related and 12 quality-related bad practices.

\item We propose \textsc{SCALM}, an LLM-based framework for smart contract bad practices auditing. This framework integrates context-aware function-level slicing and multi-layer reasoning verification (syntax, design patterns, architecture) to generate structured audit reports.

\item We conduct comprehensive experiments across multiple datasets and LLMs. Results demonstrate that \textsc{SCALM} performs well and outperforms existing tools for smart contract bad practice detection.
At the same time, ablation experiments reveal that RAG and multi-layer reasoning verification can improve \textsc{SCALM} performance.

\item We open source \textsc{SCALM}'s codes and experimental data at \url{https://doi.org/10.6084/m9.figshare.28008167}.
\end{itemize}

\section{Background}
\subsection{Large Language Models}
LLMs are trained using deep learning techniques to understand and generate human language. They are typically based on the Transformer architecture, such as ChatGPT \cite{Kasneci_2023_ChatGPT}, BERT \cite{Devlin_2019_BERT}, and GLM \cite{Du_2022_GLM}.
The training process usually involves learning language patterns and structures from large-scale text corpora. These corpora can include a variety of texts such as news articles, books, web pages, and other forms of human linguistic expression. These model can generate coherent and meaningful text by learning from these corpora. 
Furthermore, LLMs can handle various natural language processing tasks, including text generation, text classification, sentiment analysis, question-answering systems, etc \cite{10.1145/3709358,li2025ckg,li2025facial}.

One of the key features of LLMs is their powerful generative ability. These models can generate new, coherent text similar in grammar and semantics to the training data \cite{wu2023autogen}. This makes LLMs useful for various applications, including machine translation, text summarization, sentiment analysis, dialogue systems, and other natural language processing tasks \cite{10.1145/3699597}. Another important feature of LLMs is their "zero-shot" capability, which allows them to perform various tasks without any task-specific training \cite{abbasiantaeb2024let}. For example, the model can choose the most appropriate answer given a question and some answer options. This ability makes LLMs very useful in many practical applications.

However, LLMs also have some challenges and limitations. For instance, they may generate inaccurate or misleading information and reflect biases in the training data \cite{YAO2024100211}. 
LLM can adopt the RAG to improve quality and accuracy. This method combines pre-trained parametric models with non-parametric memory to enhance the quality and accuracy of smart contract code audits \cite{Gao_2024_Retrieval-Augmented}. The RAG integrates the processes of retrieval and generation into one. 

\begin{figure}[tbp]
\centering
\includegraphics[width=0.6\linewidth]{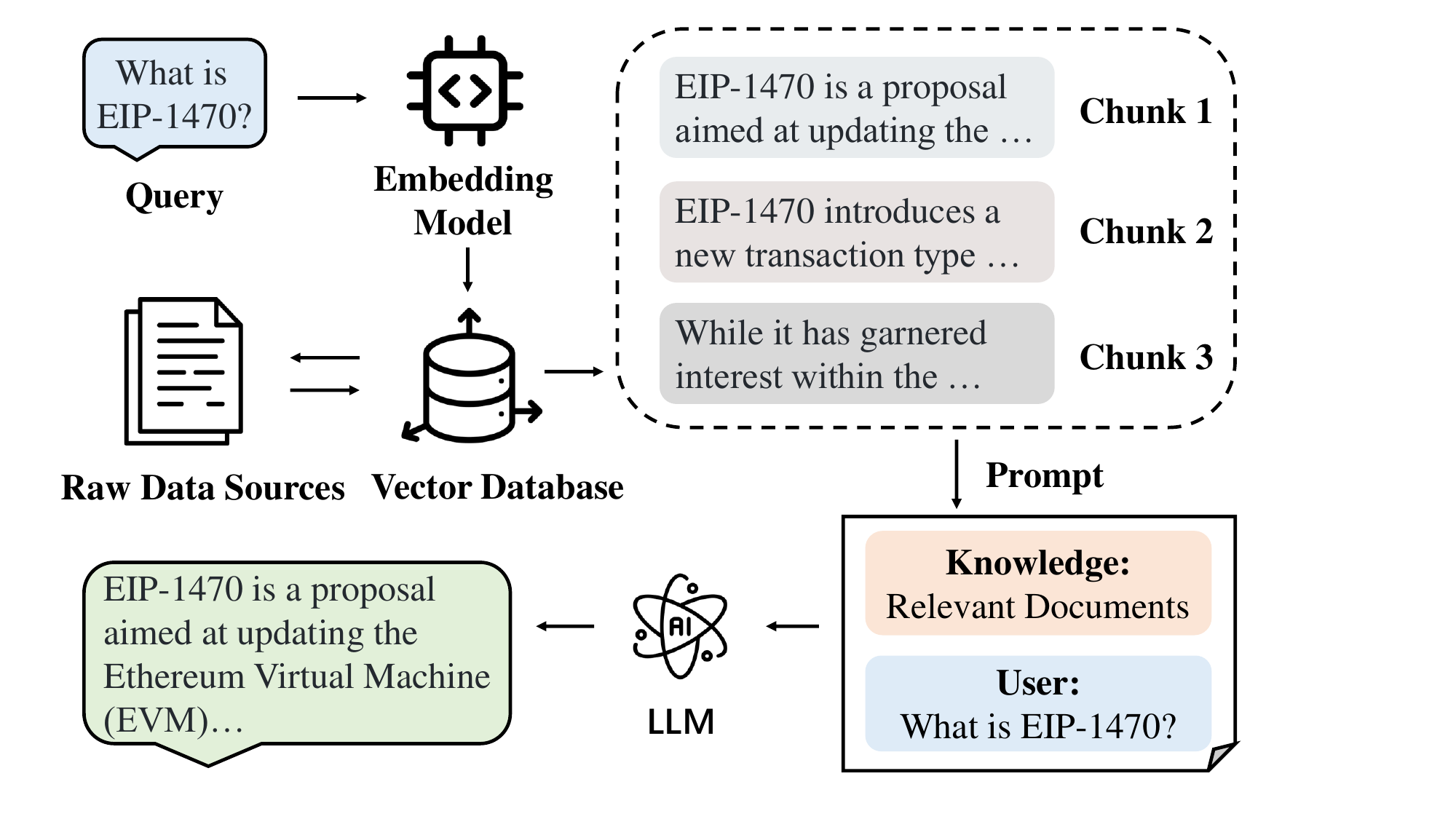}
\vspace{-0.5em}
\caption{Illustration of the RAG framework. The system processes user queries through an embedding model, retrieves relevant document chunks from a vector database, and combines the retrieved knowledge with the original query to generate contextually-enhanced responses via LLM.}
\label{fig:rag}
\vspace{-1.5em}
\end{figure}

As Fig. \ref{fig:rag} illustrates, during the operation of the model, it first retrieves relevant documents or entities from a large-scale knowledge base. Then, it inputs this retrieved information as additional context into the generation model, which generates corresponding outputs based on these inputs. This design allows RAG to utilize external knowledge bases effectively while demonstrating excellent performance when dealing with tasks requiring extensive background knowledge \cite{dos2024domain}.

\subsection{Smart Contract Weakness Classification}  

Smart contracts are self-executing protocols that run on the blockchain and allow trusted transactions without third-party intervention \cite{gao2025implementation,liu2025empirical}. However, since their code is publicly available and cannot be changed once deployed, the security of smart contracts has become an important issue \cite{10123449}. To address this issue, EIP-1470 \cite{wagner2018eip} proposes the Smart Contract Weakness Classification (SWC), a classification scheme designed to help developers identify and prevent smart contract weaknesses.

SWC concerns weaknesses that can be identified within a smart contract's Solidity code \cite{soud2024fly,li2025uscsa,li2025atomgraph}. It is designed to reference the structure and terminology of the Common Weakness Enumeration (CWE) but adds several weakness classifications specific to smart contracts \cite{chen2024healthier}. These classifications include but are not limited to, reentry attacks, arithmetic overflow, and delegatecall to untrusted callee. The example in Fig.~\ref{fig:SWC112} illustrates how SWC guides secure design: the vulnerable \texttt{Proxy} contract allows any caller to invoke \texttt{delegatecall} with arbitrary addresses, while the fixed version restricts the \texttt{callee} address to an owner-controlled parameter, mitigating unauthorized code execution risks.

All work on SWC has been incorporated into the EEA EthTrust Security Level Specification, a specification proposed by the Enterprise Ethereum Alliance (EEA) to provide a reliable methodology for assessing the security of smart contracts. 

\begin{figure}[tbp]
\begin{lstlisting}[numbers=none]
contract Proxy {
  address owner;
  constructor() public {
    owner = msg.sender;  
  }
  function forward(address callee, bytes _data) public {
    require(callee.delegatecall(_data)); // Allows ANY caller to execute delegatecall to ANY contract
  }
}

contract Proxy_fixed {
  address callee;
  address owner;
  modifier onlyOwner {
    require(msg.sender == owner);
    _;
  }
  constructor() public {
    callee = address(0x0);
    owner = msg.sender;
  }
  function setCallee(address newCallee) public onlyOwner {
    callee = newCallee; // Owner-controlled target update
  }
  function forward(bytes _data) public {
    require(callee.delegatecall(_data)); // Uses preconfigured callee address
  }
}
\end{lstlisting}
\caption{SWC-112: Delegatecall to Untrusted Callee vulnerability example. The vulnerable Proxy contract allows arbitrary delegatecall execution, while Proxy\_fixed implements access control and trusted callee validation to mitigate the security risk.}
\label{fig:SWC112}
\end{figure}

\section{Method}

\begin{figure}[tbp]
\centering
\includegraphics[width=\linewidth]{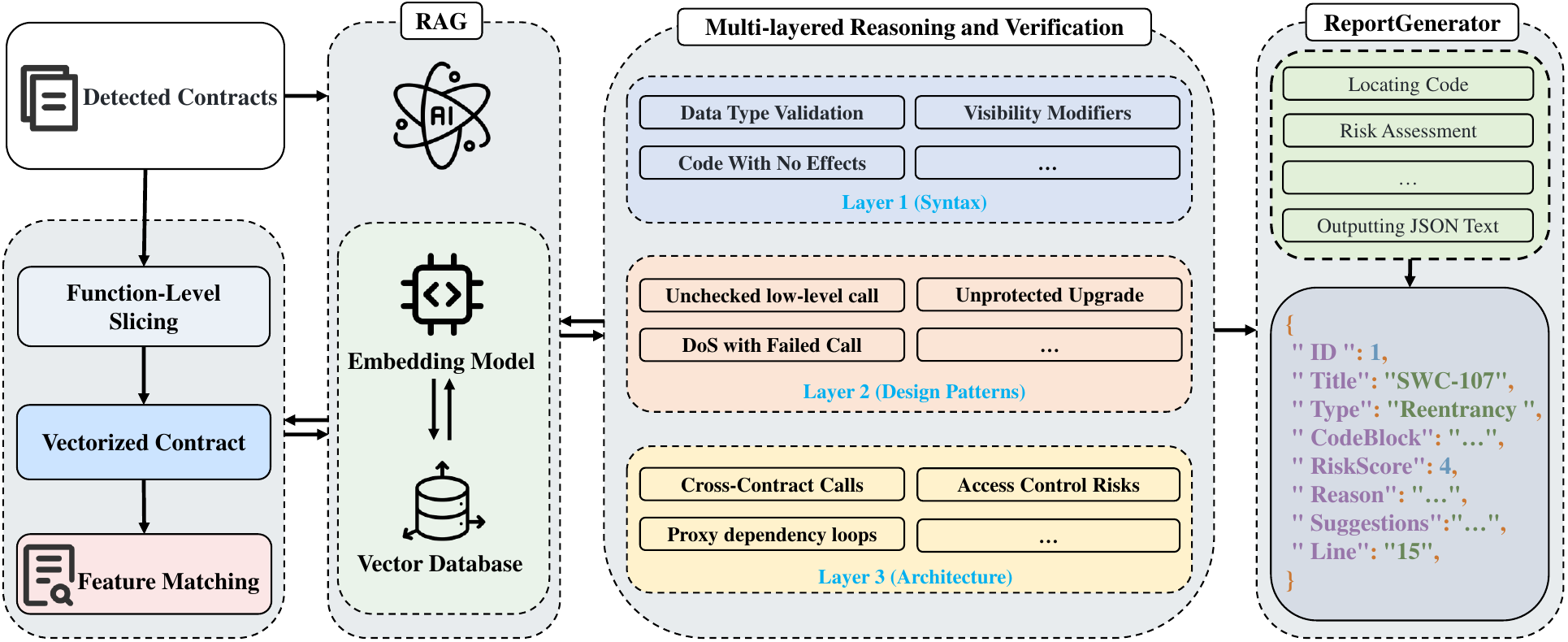}
\caption{Overall Architecture of SCALM. The framework comprises context-aware function-level slicing, contract vectorization, RAG-based retrieval, multi-layered reasoning verification (syntax, design patterns, and architecture), and automated report generation with structured JSON output.}
\label{fig:llm}
\end{figure}

Fig. \ref{fig:llm} shows the architecture of \textsc{SCALM}, which establishes a systematic framework for detecting bad practices in smart contracts through multi-layer reasoning and verification \cite{sun2025FIRESmartContract,heCodeNotNatural2024}. \textsc{SCALM} consists of two core modules \underline{: (1)} To statically analyze and extract bad practice patterns, convert them into semantic vectors, and construct an extensible knowledge base. \underline{(2)} Combines RAG and Step-Back prompting in a multi-layer reasoning verification system, which realizes multi-stage verification through layered abstraction from code syntax to architecture. Ultimately, \textsc{SCALM} leverages LLMs to generate structured audit reports. 

\subsection{Context-Aware Function-Level Slicing}

We propose a hierarchical function-level slicing approach that preserves semantic context beyond isolated code units. Unlike traditional program slicing that focuses solely on data/control flow dependencies, our method constructs \textit{context-enriched function slices} that capture both intra-function logic and inter-function call semantics.

\begin{figure}[tbp]
\centering
\includegraphics[width=\linewidth]{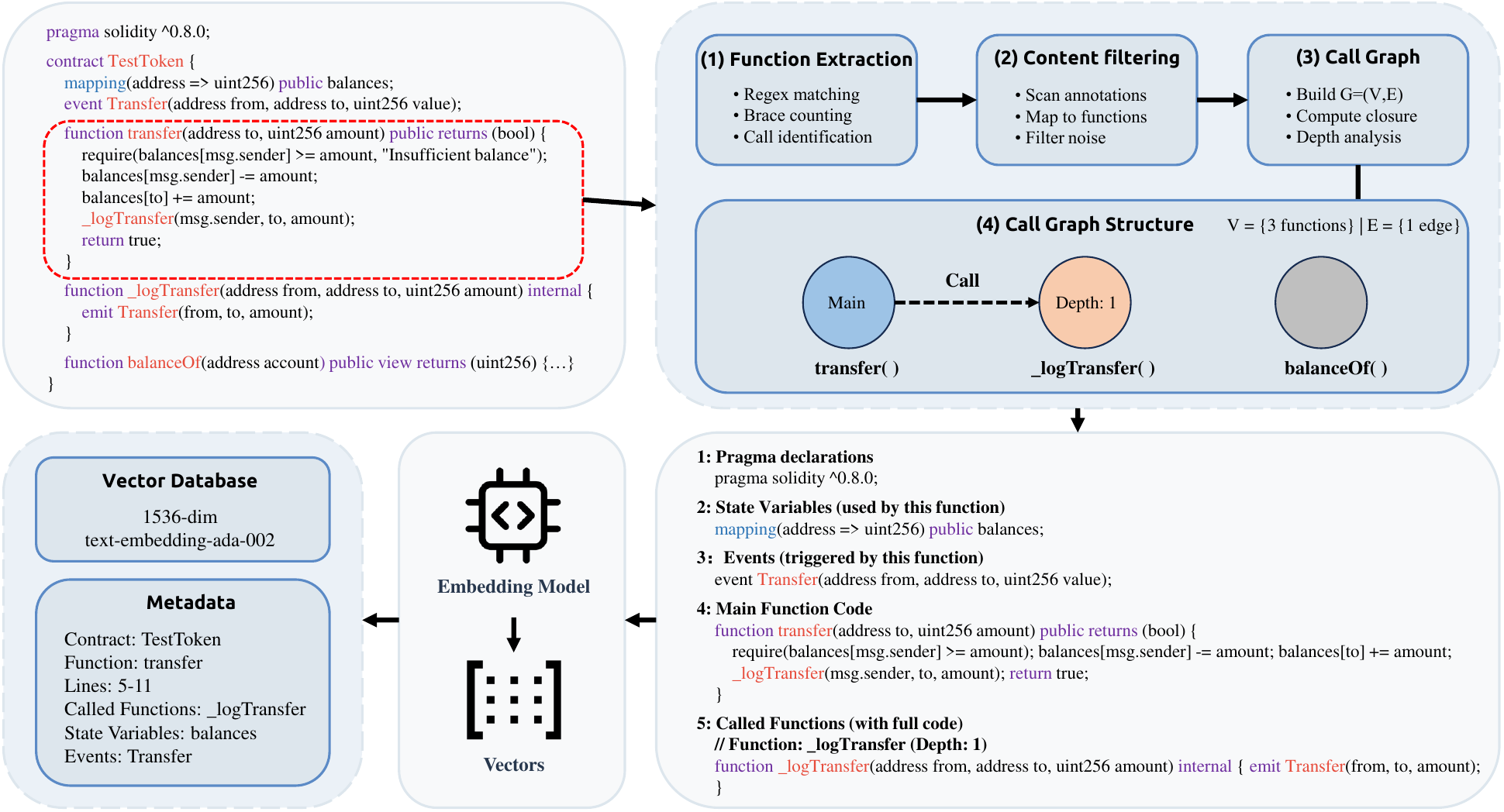}
\caption{Function-Level Context Slicing and Vectorization Pipeline. Illustration of our workflow using the \texttt{TestToken} contract as an example. The process begins by extracting a target function (\texttt{transfer}) and analyzing its dependencies (e.g., its call to \texttt{\_logTransfer}). A context-enriched slice is then assembled, incorporating the main function, its dependencies, and relevant contract-level definitions (state variables, events). Finally, this slice is vectorized and stored in a database with structured metadata for retrieval.}
\label{fig:chunk}
\vspace{-1em}
\end{figure}

Our pipeline illustrated in Fig.~\ref{fig:chunk}, operates through three principal stages:

\textbf{Function-level Decomposition with Annotation Filtering.} We first decompose the smart contract into independent functions. This is achieved through source code parsing, employing regular expression matching and brace counting techniques to identify function boundaries while preserving relevant documentation precisely. Subsequently, we employ selective extraction for knowledge base construction, focusing on bad practice-related patterns. This step is guided by SWC annotations embedded in the source code (e.g., // SWC-107: L15-20). By mapping the line numbers of these annotations to function boundaries, we construct a filtered corpus containing only functions documented with bad practices, thereby significantly improving the signal-to-noise ratio.

\textbf{Dependency-Aware Context Assembly.} For each target function $f_{\text{main}}$, we construct an enriched context $C(f_{\text{main}})$ that captures its semantic environment. This process begins with building a contract-wide directed call graph $G=(V,E)$ to analyze inter-function dependencies. The enriched context is then assembled as follows:
\begin{equation}
C(f_{\text{main}}) = \{\text{pragmas}\} \cup V_{\text{rel}} \cup E_{\text{rel}} \cup \{f_{\text{main}}\} \cup D(f_{\text{main}}, d_{\max})
\end{equation}
where $V_{\text{rel}}$ represents state variables referenced within $f_{\text{main}}$, $E_{\text{rel}}$ denotes triggered event definitions, and $D(f, d)$ recursively includes called functions up to a depth of $d_{\max} = 3$. This bounded recursion is formalized as:
\begin{equation}
D(f, d) = \begin{cases}
\emptyset & \text{if } d = 0 \\
\bigcup_{g \in \text{calls}(f)} \{g\} \cup D(g, d-1) & \text{otherwise}
\end{cases}
\end{equation}
Fig.~\ref{fig:chunk} illustrates this process with the \texttt{TestToken} contract, the assembled slice is a multi-layered composition including: (1) pragma declarations, (2) relevant state variables (e.g., \texttt{balances}), (3) relevant event definitions (e.g., \texttt{Transfer}), (4) the main function’s code, and (5) the full code of called functions (e.g., \texttt{\_logTransfer()} at depth 1). The depth limit balances capturing sufficient context while avoiding noise from deep call chains that could reduce embedding quality.

\textbf{Structured Metadata Augmentation.} Each context slice is annotated with structured metadata, including source information (source file, contract name), bad practice classification (extracted SWC types), and dependency details (e.g., called functions, referenced state variables, and triggered events). This design, which combines vectorised code content with its corresponding structured metadata, forms a dual representation enabling semantic similarity retrieval and fine-grained filtering during multi-layered reasoning and verification.

\subsection{Semantic Vectorization and Retrieval}
We use a vector database to store and query large amounts of vector data. 
A vector database is a particular database that can store and query large amounts of vector data \cite{Hambardzumyan_2022_Deepa}. In the vector database, data is stored as vectors, each typically represented by a set of floating-point numbers. These vectors can represent various data types, such as images, audio, text, etc. In bad practice detection tasks, through the embedding model (i.e., \texttt{text-embedding-ada-002}), code slices are transformed into 1536-dimensional vectors that preserve syntactic structures and semantic relationships. This process captures both explicit bad practice patterns and implicit code quality issues. This process can be expressed with the following \cref{eqa:embedding}:

\begin{equation}
\begin{aligned}
    \vec{v} = f_{\text{Embedding}}(\text{Text})
\label{eqa:embedding}
\end{aligned}
\end{equation}

Where $f_{\text{Embedding}}$ is our embedding model, $\text{Text}$ is the input text, and $\vec{v}$ is the outputted vector.
This vectorized data storage method significantly improves efficiency in handling it. Firstly, storing data as vectors makes it more compact, thus reducing storage space requirements. Secondly, vectorized data facilitates parallel computing, which is crucial when dealing with large-scale datasets. A vital feature of a Vector Database lies in its ability to perform efficient similarity searches, which are notably advantageous when dealing with high-dimensional datasets. This similarity search can be achieved by calculating cosine similarities between two vectors  with the \cref{eqa:similar}:

\begin{equation}
\begin{aligned}
    \text{similarity}(A,B)= \frac {A \cdot B}{||A||_2 \cdot ||B||_2}
\label{eqa:similar}
\end{aligned}
\end{equation}

where $A$ and $B$ are two vectors, $A \cdot B$ is their dot product, and $||A|||_2$ and $||B|||_2$ are their L2-norm (Euclidean norm).

The system establishes logical constraint verification through vector similarity retrieval to trigger SWC-standard-based semantic reasoning. For all code snippets $c \in \text{CodeSnippets}$, if there exists a rule $r \in \text{SWC-Rules}$ where the semantic match score $\text{Match}(v(c),v(r))$ exceeds threshold $\theta$, the code snippet is flagged as suspicious \cref{eqa:match}: 

\begin{equation}
\begin{aligned}
    \forall c \in \text{CodeSnippets}, \exists r \in \text{VDB-Rules}, \text{Match}(v(c),v(r)) > \theta \Rightarrow \text{Flag}(c)
\label{eqa:match}
\end{aligned}
\end{equation}

Here, $v(c)$ and $v(r)$ represent the vector embeddings of code snippets and vector database (VDB) rules, respectively, generated through the same embedding model. The matching operation employs the cosine similarity defined in \cref{eqa:similar}, with threshold $\theta$ = 0.9 empirically calibrated to balance precision and recall.


\subsubsection{Dynamic Pattern Updating}
 When encountering code patterns not present in the current knowledge base (i.e., vector similarity searches yield no matches above predefined confidence thresholds), the system initiates multi-layered reasoning and verification processes to analyze potential bad practices. Following confirmation through multi-layered reasoning and verification, newly identified vulnerabilities undergo automated annotation with SWC identifiers and semantic vectorization using the embedding model.

\subsection{Multi-layer Reasoning and Verification}


\textbf{Workflow Overview.} The verification process operates on individual function slices obtained from context-aware function-level slicing. Each function slice is processed independently through the following pipeline: (1) RAG-based retrieval identifies relevant bad practice patterns from the vector database for the target function. (2) The function undergoes three sequential layers of verification (Syntax, Design Patterns, Architecture), where each layer employs Step-Back prompting to connect code-level patterns with security principles. (3) Individual function-level audit results are collected and aggregated into a comprehensive contract-wide report. This function-by-function analysis ensures that each code unit is evaluated with complete contextual information while maintaining independent verification across different functions.

We employ Step-Back prompting \cite{Zheng__TAKE} to bridge concrete code patterns with abstract security principles. This technique uses the capabilities of LLMs to abstract high-level concepts and basic principles from specific code instances. In this way, not only can the model understand the literal meaning of the code, but it can also comprehend underlying logic and potential design patterns through abstract thinking. Step-Back prompting consists of two main steps:
\begin{itemize} 
\item Abstraction: Instead of directly posing questions, we propose a step-back question about higher-level concepts or principles and retrieve facts related to these higher-level concepts or principles. In detecting bad practices in smart contracts, we use abstract prompts that aim to guide LLMs to explore the literal meaning of code and deeper structures and intentions. These prompts may include questions like "What are the potential risks with this implementation?" or "Does this method comply with basic principles for secure smart contracts?"

\item Model Reasoning: Based on facts about high-level concepts or principles, LLMs can reason about answers to the original question. We refer to this as abstraction-based reasoning. Reasoning with these abstraction hints attempts to analyze the code from a broader perspective. This includes comparing the strengths and weaknesses of different implementations and how they fit with known best practices or common bad practices.
\end{itemize}


For instance, when detecting the SWC-112  (Delegatecall to Untrusted Callee), the model first considers the underlying mechanism by asking questions such as "Based on the fundamental principles of smart contract security, does the code of smart contract contain any bad practices?". The model then maps this principle to the code implementation level. It checks whether the contract logic strictly authenticates the call target of the delegatecall function.
If the whitelisting mechanism does not constrain the target address or if there is a risk of dynamic injection, the model determines that the code violates the security principle and identifies the bad practice. 


\begin{figure}[tbp]
\centering
\includegraphics[width=\linewidth]{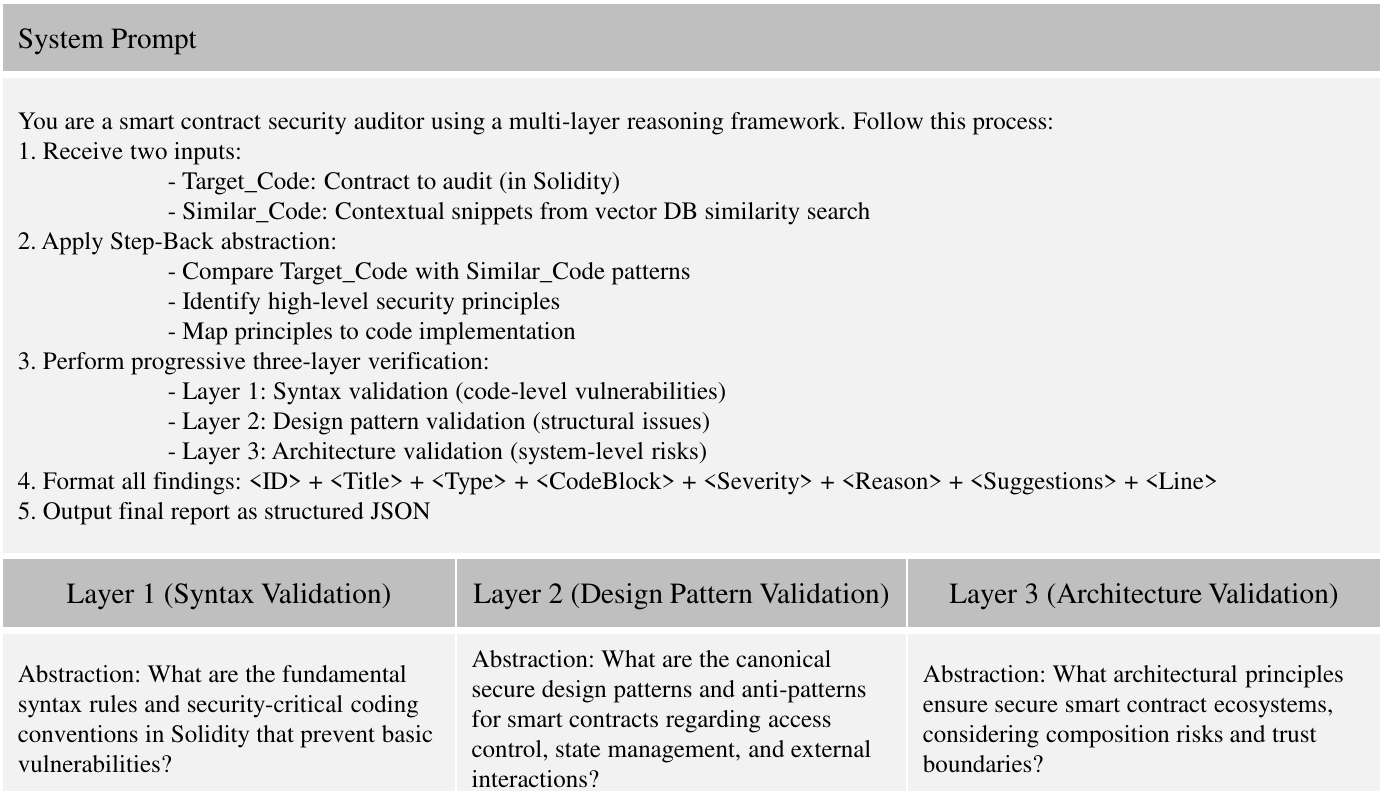}
\caption{The prompt template used by roles. The system prompt defines a multi-layer reasoning framework with step-back prompting. Each layer (Syntax, Design Pattern, and Architecture) has specialized abstractions to guide LLM analysis from code-level vulnerabilities to system-level architectural risks, ensuring comprehensive security auditing.}
\label{fig:prompt}
\end{figure}

\subsubsection{Compliance Validation Pipeline}
As shown in Fig.~\ref{fig:prompt}, the validation process implements a three-layer verification hierarchy that mirrors secure development lifecycles. For each function slice, the system performs three reasoning passes, with each layer leveraging Step-Back prompting to elevate the analysis from code to abstract security principles:

\begin{itemize}
\item \textbf{Layer 1 (Syntax)}: Static rule checking using LLM-powered semantic pattern recognition, validating compliance with Solidity-specific constraints (e.g., Delegatecall to Untrusted Callee, visibility modifiers). Step-Back prompting guides the LLM to first consider "What are the fundamental syntax-level security requirements?" before examining specific code constructs.

\item \textbf{Layer 2 (Design Patterns)}: LLM-powered analysis of design pattern implementation. Verify correct application of common Solidity design patterns (e.g., Checks-Effects-Interactions, Access Control patterns) and identify anti-patterns that may lead to vulnerabilities. Step-Back prompting asks "What design principles should this pattern follow?" to bridge pattern recognition with architectural intent.

\item \textbf{Layer 3 (Architecture)}: LLM-driven contract architecture analysis. Evaluate contract architectural risks and verify compliance with best practices by parsing the topology and invocation relationships of smart contracts. Step-Back prompting elevates analysis to "What are the system-level security implications?" connecting individual functions to ovethe rall contract architecture.
\end{itemize}

The multi-layer reasoning validation results are comprehensively evaluated through the dynamic weighted scoring \cref{eqa:risk}:


\begin{equation}
\begin{aligned}
\text{RiskScore} = \frac{1}{3} \sum_{i=1}^3 \text{Severity}_i
\label{eqa:risk}
\end{aligned}
\end{equation}

The formula given by \cref{eqa:risk} calculates a comprehensive risk score by averaging findings across three distinct validation stages ($i=1$ to $3$), each corresponding to a specific security assessment layer.
 For each layer, the detected bad practices are assigned a normalized severity value derived from CVSS v3.1 baselines \cite{cvss}.

\subsubsection{Audit Report Generation}

After each function slice completes the three-layer verification process, the LLM generates a function-level audit report documenting identified bad practices. These individual reports are then aggregated into a comprehensive contract-wide audit document. The report is output in \texttt{JSON} format. It contains the bad practice \textit{ID},  \textit{Title}, \textit{Type}, specific \textit{CodeBlock} along with its \textit{Location}, \textit{RiskScore},  \textit{Reason} for the problem, and \textit{Suggestions} for improvement. This aggregation process consolidates findings across all functions while preserving the detailed analysis from each verification layer, enabling developers to understand both function-specific issues and contract-wide bad practice patterns.

Through this method, we can effectively utilize the powerful capabilities of LLMs for deep security audits on smart contracts, thereby helping developers identify and fix potential security issues. The Algorithm \ref{alg:scalm} outlines the step-by-step procedure for generating an audit report for smart contract code. 

\setlength{\textfloatsep}{5pt}
\begin{algorithm}[htbp]
\caption{Muti-layer Reasoning Verification with RAG}
\begin{algorithmic}[1]
\Procedure{DataProcessing}{$\mathcal{C}$} \Comment{Smart contract set}
    \State $\mathcal{S} \gets \text{FunctionLevelSlicing}(\mathcal{C})$
    \State $\mathcal{P} \gets \text{ExtractBadPractices}(\mathcal{S})$
    \State $\mathcal{V} \gets \text{SemanticVectorization}(\mathcal{P})$ \Comment{Eq.~\ref{eqa:embedding}}
    \State $\text{UpdateVectorDB}(\mathcal{V})$
\EndProcedure
\Procedure{MultiLayerVerification}{$c$} \Comment{Target contract}
    \State $\mathcal{F} \gets \text{FunctionLevelSlicing}(c)$ \Comment{Extract function slices}
    \State $\text{Reports} \gets \emptyset$
    \For{each function slice $f \in \mathcal{F}$}
        \State $\mathcal{M} \gets \text{RAGRetrieval}(\mathcal{V}, f)$ \Comment{Eq.~\ref{eqa:similar}}
        \State $\text{Layer1Check}(\mathcal{M}, f)$ \Comment{Syntax validation}
        \State $\text{Layer2Check}(\mathcal{M}, f)$ \Comment{Design patterns}
        \State $\text{Layer3Check}(\mathcal{M}, f)$ \Comment{Architecture analysis}
        \State $\text{RiskScore}_f \gets \frac{1}{3} \sum_{i=1}^3 \text{Severity}_i$ \Comment{Eq.~\ref{eqa:risk}}
        \State $\text{Reports} \gets \text{Reports} \cup \text{GenerateFunctionReport}(f)$
    \EndFor
    \State \Return $\text{AggregateReports}(\text{Reports})$ \Comment{Contract-wide report}
\EndProcedure
\Function{RAGRetrieval}{$\mathcal{V}, c$}
    \State $\vec{v}_c \gets f_{\text{Embedding}}(c)$
    \State \Return $\{r \in \mathcal{V} \mid \text{sim}(\vec{v}_c, \vec{v}_r) > \theta\}$
\EndFunction
\Function{DynamicUpdate}{$\mathcal{P}_{\text{new}}$}
    \If{$\nexists r \in \mathcal{V} \mid \text{sim}(\mathcal{P}_{\text{new}}, r) > \theta$}
        \State $\text{MultiLayerVerify}(\mathcal{P}_{\text{new}})$
        \State $\text{UpdateVectorDB}(\mathcal{P}_{\text{new}})$
    \EndIf
\EndFunction
\end{algorithmic}
\label{alg:scalm}
\end{algorithm}

\section{Experiments}
\subsection{Experiment Settings}
All experiments are executed on a server equipped with NVIDIA GeForce GTX 4070Ti GPU, Intel(R) Core(TM) i9-13900KF CPU, and 128G RAM, operating on Ubuntu 22.04 LTS. The software environment includes Python 3.9 and PyTorch 2.0.1.

\textbf{Dataset.}
In this paper, we use the \textbf{DAppSCAN} dataset \cite{zheng2023dappscan} as a knowledge base for detecting bad practices in smart contracts. The dataset contains 39,904 Solidity files with 1,618 SWC weaknesses from 682 real projects. The \textbf{Smartbugs} dataset \cite{DurieuxEtAl2020ICSE} is also used in the experiment, and a total of 1,894 smart contracts with five types of security-related bad practices (SWC weaknesses) are extracted for comparison experiments. Additionally, we introduce the \textbf{SolQuality} dataset \cite{solquality}, comprising 1,200 smart contracts covering six categories of quality-related bad practices. This dataset encompasses 25 diverse contract scenarios, including ERC20 tokens, NFTs \cite{li2025beyond}, voting systems, auctions, and other common decentralized application patterns \cite{long2025fomo3d,sun2025data}, providing comprehensive coverage for quality-oriented bad practices detection. These datasets form the basis of our experimental analysis. Table \ref{table:dataset} summarizes the smart contract data used.

\textbf{Models.}
For the selection of LLMs, we chose six current state-of-the-art models for detection experiments. GPT-4o, GPT-4-1106-preview, and GPT-4-0409 are the latest versions from OpenAI with powerful natural language processing capabilities \cite{achiam2023gpt}. Claude-3.5-Sonnet is Anthropic's new-generation model focused on safety and interoperability \cite{jiang2025deepseek}. Gemini-1.5-Pro is Google's high-performance model optimized for multitasking \cite{team2024gemini}. Llama-3.1-70b-Instruct is Meta's large-scale model specializing in instruction following and generating high-quality text \cite{grattafiori2024llama}.

\begin{table}[htbp]
\centering
\caption{The Collected Dataset for Our Evaluation. \# indicates the number of each item.}
\begin{tabular}{>{\centering\arraybackslash}p{2.8cm} c c}
\toprule
\textbf{Dataset} & \textbf{\# Contracts} & \textbf{Purpose} \\
\midrule
DAppSCAN & 39,904 & Knowledge Base \\
\midrule
Smartbugs & 1,894 & Security-related \\
\midrule
SolQuality & 1,200 & Quality-related \\
\bottomrule
\end{tabular}

\label{table:dataset}
\vspace{-1em}
\end{table}

\noindent\textbf{Evaluation Metrics.} We treat the evaluation of detection accuracy as a binary classification task, where each contract is classified as either containing or not containing a specific bad practice. We carry out experiments to answer the following research questions:\\
\underline{\textbf{RQ1:}} How robust is the RAG component to code mutations? What is the impact of the similarity threshold on retrieval performance?\\
\underline{\textbf{RQ2:}} How effective is \textsc{SCALM} in detecting bad practices in smart contracts? How do different LLMs affect \textsc{SCALM}?\\
\underline{\textbf{RQ3:}}  Can \textsc{SCALM} find bad practices undetectable by other tools? How does it compare with existing tools?\\
\underline{\textbf{RQ4:}} How do RAG and multi-layer reasoning components contribute to \textsc{SCALM}'s detection performance?


\subsection{RQ1: RAG Robustness and Retrieval Performance}
The effectiveness of RAG-based systems heavily depends on the quality and robustness of the retrieval component. To ensure that \textsc{SCALM}'s RAG module provides reliable code retrieval under various practical scenarios, we conduct comprehensive experiments to evaluate its robustness. Specifically, we investigate two critical aspects: (1) robustness to code mutations, and (2) sensitivity to similarity threshold.

\textbf{Robustness to Code Mutations.} We evaluate how well the RAG system retrieves relevant code samples under various code mutations commonly occurring in real-world scenarios. Table~\ref{tab:rag_robustness} shows the retrieval performance across different mutation types. We test 50 randomly sampled functions from the vector database with five mutation types: (1) Variable Rename: 70\% of variables renamed. (2) Dead Code: three unreachable code blocks inserted. (3) Comment Add: five comments added. (4) Comment Remove: 80\% of comments removed. (5) Combined: applying multiple mutations simultaneously. We perform RAG retrieval with top-10 results for each mutated sample and measure whether the original sample is retrieved within top-K ranks. 

The results demonstrate that the RAG system maintains strong robustness, with Recall@3 consistently above 92\% across all mutation types. Notably, the system achieves 100\% Recall@3 for dead code insertion and comment addition, indicating excellent resilience to these transformations. The Mean Reciprocal Rank (MRR) values range from 0.7585 to 0.8609, showing that relevant code samples are typically ranked high in the retrieval results. The combined mutation scenario demonstrates that the RAG system maintains reliable retrieval performance even under multiple simultaneous transformations.

\begin{table}[tbp]
  \centering
  \caption{RAG System Robustness under Code Mutations}
  \label{tab:rag_robustness}
  \begin{tabular}{lcccccc}
  \toprule
  Mutation Type & Recall@1 (\%) & Recall@3 (\%) & Recall@5 (\%) & Recall@10 (\%) & MRR \\
  \midrule
  Variable Rename & 58.00 & 92.00 & 96.00 & 98.00 & 0.7585 \\
  Dead Code & 66.00 & 100.00 & 100.00 & 100.00 & 0.8100 \\
  Comment Add & 74.00 & 100.00 & 100.00 & 100.00 & 0.8600 \\
  Comment Remove & 74.00 & 94.00 & 96.00 & 98.00 & 0.8609 \\
  Combined & 68.00 & 94.00 & 96.00 & 98.00 & 0.8206 \\
  \bottomrule
  \end{tabular}
\end{table}

\begin{table}[htbp]
\centering
\caption{Threshold Sensitivity Analysis}
\label{tab:threshold_sensitivity}
\begin{tabular}{cccccc}
\toprule
Threshold & Recall@1 (\%) & Recall@3 (\%) & Recall@5 (\%) & Recall@10 (\%) & Retention Rate (\%) \\
\midrule
0.70 & 53.33 & 90.00 & 93.33 & 96.67 & 100.00 \\
0.80 & 53.33 & 90.00 & 93.33 & 96.67 & 100.00 \\
0.85 & 53.33 & 90.00 & 93.33 & 96.67 & 100.00 \\
0.90 & 50.00 & 83.33 & 86.67 & 90.00 & 56.67 \\
0.95 & 3.33 & 6.67 & 6.67 & 6.67 & 1.67 \\
\bottomrule
\end{tabular}
\end{table}

\textbf{Similarity Threshold Analysis.} Table~\ref{tab:threshold_sensitivity} presents the impact of different similarity thresholds on retrieval performance. The results show that thresholds from 0.70 to 0.85 maintain consistent retrieval performance with 100\% retention rate, while the 0.90 threshold provides a balanced trade-off between precision and recall. At threshold 0.90, the system achieves 50.00\% Recall@1 and 83.33\% Recall@3, with a retention rate of 56.67\%, effectively filtering out dissimilar code while maintaining good recall. In contrast, the 0.95 threshold is too restrictive, with only 1.67\% retention rate and dramatically reduced recall performance. Our analysis reveals that the similarity distribution has an average of 0.9041 ± 0.0200, with a median of 0.9029 and a 75th percentile of 0.9159. Notably, 276/500 (55.2\%) of retrieval results have similarity $\geq$ 0.9, while only 11/500 (2.2\%) exceed 0.95. These statistics validate that the 0.90 threshold is well-calibrated for our task, as it effectively distinguishes between relevant and irrelevant code pairs without being overly restrictive.

\begin{tcolorbox}[boxrule=1pt,boxsep=1pt,left=2pt,right=2pt,top=1pt,bottom=1pt]
\textbf{Answer to RQ1.} The RAG module in \textsc{SCALM} demonstrates strong robustness and reliable retrieval performance. Across various code mutations, the RAG module's Recall@3 consistently remains above 92\%. A similarity threshold of 0.90 provides the optimal balance, achieving 83.33\% Recall@3.
\end{tcolorbox}



\subsection{RQ2: Bad Practice Detection}

\begin{table*}[tb]
\centering
\caption{SWC bad practice detection. Full model names are GPT-4o, GPT-4-1106-preview, GPT-4-0409, Claude-3.5-Sonnet, Gemini-1.5-Pro, and Llama-3.1-70b-Instruct. A checkmark (\ding{51}) indicates successful detection, and a cross (\ding{55}) indicates a failure.}
\resizebox{\textwidth}{!}{
\begin{tabular}{@{}lccccccc@{}}
\toprule
\multirow{2}{*}{\textbf{Category}} & \multirow{2}{*}{\textbf{Title}} & \multicolumn{6}{c}{\textbf{Models}} \\
\cline{3-8}
 & & GPT-4o & GPT-4-1106 & GPT-4-0409 & Claude & Gemini & Llama \\
\midrule
SWC-100 & Function Default Visibility & \ding{51} & \ding{55} & \ding{55} & \ding{55} & \ding{55} & \ding{55} \\
SWC-101 & Integer Overflow and Underflow & \ding{51} & \ding{51} & \ding{51} & \ding{51} & \ding{51} & \ding{51} \\
SWC-102 & Outdated Compiler Version & \ding{51} & \ding{51} & \ding{51} & \ding{51} & \ding{51} & \ding{51} \\
SWC-103 & Floating Pragma & \ding{51} & \ding{51} & \ding{51} & \ding{51} & \ding{51} & \ding{51} \\
SWC-104 & Unchecked Call Return Value & \ding{51} & \ding{51} & \ding{51} & \ding{51} & \ding{51} & \ding{51} \\
SWC-105 & Unprotected Ether Withdrawal & \ding{51} & \ding{51} & \ding{51} & \ding{51} & \ding{51} & \ding{51} \\
SWC-106 & Unprotected SELFDESTRUCT Instruction & \ding{51} & \ding{51} & \ding{51} & \ding{51} & \ding{51} & \ding{51} \\
SWC-107 & Reentrancy & \ding{51} & \ding{51} & \ding{51} & \ding{51} & \ding{51} & \ding{51} \\
SWC-108 & State Variable Default Visibility & \ding{51} & \ding{51} & \ding{51} & \ding{51} & \ding{51} & \ding{51} \\
SWC-109 & Uninitialized Storage Pointer & \ding{55} & \ding{55} & \ding{55} & \ding{55} & \ding{55} & \ding{55} \\
SWC-110 & Assert Violation & \ding{51} & \ding{51} & \ding{51} & \ding{51} & \ding{51} & \ding{51} \\
SWC-111 & Use of Deprecated Solidity Functions & \ding{51} & \ding{51} & \ding{51} & \ding{51} & \ding{51} & \ding{51} \\
SWC-112 & Delegatecall to Untrusted Callee & \ding{51} & \ding{51} & \ding{51} & \ding{51} & \ding{55} & \ding{55} \\
SWC-113 & DoS with Failed Call & \ding{51} & \ding{51} & \ding{51} & \ding{51} & \ding{55} & \ding{51} \\
SWC-114 & Transaction Order Dependence & \ding{55} & \ding{55} & \ding{55} & \ding{55} & \ding{51} & \ding{55} \\
SWC-115 & Authorization through tx.origin & \ding{51} & \ding{51} & \ding{51} & \ding{51} & \ding{51} & \ding{51} \\
SWC-116 & Block values as a proxy for time & \ding{51} & \ding{51} & \ding{51} & \ding{51} & \ding{51} & \ding{51} \\
SWC-117 & Signature Malleability & \ding{51} & \ding{51} & \ding{51} & \ding{55} & \ding{55} & \ding{51} \\
SWC-118 & Incorrect Constructor Name & \ding{51} & \ding{55} & \ding{51} & \ding{51} & \ding{55} & \ding{55} \\
SWC-119 & Shadowing State Variables & \ding{51} & \ding{51} & \ding{51} & \ding{51} & \ding{51} & \ding{51} \\
SWC-120 & Weak Sources of Randomness from Chain Attributes & \ding{51} & \ding{51} & \ding{51} & \ding{51} & \ding{51} & \ding{51} \\
SWC-123 & Requirement Violation & \ding{55} & \ding{51} & \ding{55} & \ding{51} & \ding{51} & \ding{51} \\
SWC-124 & Write to Arbitrary Storage Location & \ding{51} & \ding{55} & \ding{51} & \ding{55} & \ding{55} & \ding{55} \\
SWC-125 & Incorrect Inheritance Order & \ding{55} & \ding{55} & \ding{55} & \ding{55} & \ding{55} & \ding{55} \\
SWC-126 & Insufficient Gas Griefing & \ding{51} & \ding{51} & \ding{51} & \ding{55} & \ding{55} & \ding{55} \\
SWC-127 & Arbitrary Jump with Function Type Variable & \ding{51} & \ding{51} & \ding{51} & \ding{51} & \ding{51} & \ding{51} \\
SWC-128 & DoS With Block Gas Limit & \ding{51} & \ding{51} & \ding{51} & \ding{51} & \ding{51} & \ding{51} \\
SWC-129 & Typographical Error & \ding{51} & \ding{51} & \ding{51} & \ding{51} & \ding{51} & \ding{51} \\
SWC-130 & Right-To-Left-Override control character (U+202E) & \ding{55} & \ding{55} & \ding{55} & \ding{55} & \ding{51} & \ding{51} \\
SWC-131 & Presence of unused variables & \ding{51} & \ding{51} & \ding{51} & \ding{55} & \ding{51} & \ding{55} \\
SWC-132 & Unexpected Ether balance & \ding{51} & \ding{51} & \ding{51} & \ding{51} & \ding{55} & \ding{55} \\
SWC-133 & Hash Collisions With Multiple Variable Length Arguments & \ding{55} & \ding{55} & \ding{55} & \ding{55} & \ding{55} & \ding{55} \\
SWC-134 & Message call with hardcoded gas amount & \ding{51} & \ding{51} & \ding{51} & \ding{51} & \ding{51} & \ding{51} \\
SWC-135 & Code With No Effects & \ding{51} & \ding{51} & \ding{51} & \ding{51} & \ding{51} & \ding{51} \\
SWC-136 & Unencrypted Private Data On-Chain & \ding{55} & \ding{55} & \ding{55} & \ding{51} & \ding{55} & \ding{55} \\

\multirow{2}{2.5cm}{Readability}
& Unclear variable naming
& \ding{51} & \ding{51} & \ding{51} & \ding{51} & \ding{51} & \ding{51} \\

& Inconsistent function naming
& \ding{51} & \ding{51} & \ding{51} & \ding{51} & \ding{51} & \ding{55} \\

\multirow{2}{2.5cm}{Efficiency}
& Redundant state variable self-assignment
& \ding{51} & \ding{51} & \ding{51} & \ding{51} & \ding{51} & \ding{55} \\

& Unnecessary dummy variable calculation in the loop 
& \ding{55} & \ding{51} & \ding{51} & \ding{51} & \ding{55} & \ding{55} \\
\multirow{2}{2.5cm}{Data Type} 
& Using uint256 instead of bool
& \ding{51} & \ding{51} & \ding{51} & \ding{51} & \ding{51} & \ding{51} \\

& Using dynamic bytes instead of bytes32
& \ding{51} & \ding{55} & \ding{55} & \ding{51} & \ding{51} & \ding{51} \\

\multirow{2}{2.5cm}{Function Design} 
& Too many parameters
& \ding{51} & \ding{51} & \ding{51} & \ding{51} & \ding{51} & \ding{51} \\

& Meaningless return value
& \ding{55} & \ding{51} & \ding{51} & \ding{51} & \ding{55} & \ding{51} \\

\multirow{2}{2.5cm}{Events}
& Missing event emission
& \ding{51} & \ding{51} & \ding{51} & \ding{51} & \ding{51} & \ding{51} \\

& Missing indexed keyword in event parameters
& \ding{51} & \ding{51} & \ding{55} & \ding{55} & \ding{51} & \ding{55} \\

\multirow{2}{2.5cm}{Architecture} 
& Code duplication 
& \ding{51} & \ding{51} & \ding{51} & \ding{55} & \ding{51} & \ding{51} \\

& Hardcoded magic numbers
& \ding{51} & \ding{51} & \ding{55} & \ding{55} & \ding{55} & \ding{51} \\

\bottomrule
\end{tabular}
}

\label{table:swc}
\end{table*}

To assess the effectiveness of \textsc{SCALM} in detecting bad practices within smart contracts, we conduct a comprehensive evaluation using a variety of LLMs. The primary objective is to determine how well \textsc{SCALM} identifies known bad practices. We also seek to understand the impact of different LLMs on the performance of \textsc{SCALM} in detecting these bad practices.


For each model, we evaluate the framework's ability to identify instances of bad practices across two complementary dimensions correctly. The first dimension covers 35 SWC categories that include most security-related bad practices, though they do not exhaust every possible weakness in smart contracts. The second dimension comprises 12 quality-related bad practices spanning readability, efficiency, data type usage, function design, event handling, and architecture. Detection outcomes for every item are recorded as success or failure depending on whether the LLM reports the correct SWC-ID, keyword, or predefined pattern description. All assessment results are cross-validated to ensure statistical significance: each item is tested independently at least three times to eliminate random errors.
Table \ref{table:swc} summarizes the experimental results, revealing the performance differences between the evaluated models. 

\textbf{Security-related Bad Practices.} GPT-4o achieved the highest detection rate, successfully identifying vulnerabilities across 28 SWC categories, including critical issues such as SWC-101 and SWC-107. Subsequent GPT-4 iterations (1106-preview and 0409) showed marginal decreases in performance, failing to detect SWC-100. Claude-3.5-Sonnet showed lower accuracy in SWC-117 detection, while 
Gemini-1.5-Pro had difficulty detecting SWC-114. All models consistently failed to detect SWC-109 and SWC-133, indicating current limitations in LLMs' code analysis capabilities.

\textbf{Quality-related Bad Practices.} 
Most models successfully detected readability issues, data type misuse, excessive function parameters, and code duplication. However, models showed inconsistent results on more subtle issues: efficiency problems like unnecessary loop calculations challenged GPT-4o and Gemini, while event-related issues, notably missing indexed keywords, proved challenging for GPT-4-0409, Claude, and Llama. GPT-4-0409, Claude, and Gemini missed architecture issues like hardcoded magic numbers. 
These findings indicate that while LLMs excel at detecting explicit code quality violations, they struggle with subtle semantic problems that require a deeper understanding of smart contract design patterns and gas optimization principles.

The variations in detection accuracy across LLMs suggest two strategies for improving code auditing. First, since no single model comprehensively identifies all bad practices, employing multiple specialized models, each handling vulnerability categories matching their strengths, could achieve more complete coverage. Second, selecting a high-performing model and enhancing it through additional contextual information or fine-tuning offers a unified alternative. \textsc{SCALM} can leverage multi-model specialization for comprehensive detection or single-model enhancement for consistency, aiming to mitigate current limitations in LLM-based bad practice analysis.


\begin{tcolorbox}[boxrule=1pt,boxsep=1pt,left=2pt,right=2pt,top=1pt,bottom=1pt]
\textbf{Answer to RQ2.}
Our evaluation across 35 SWC categories and 12 quality dimensions demonstrates that \textsc{SCALM} effectively identifies a broad range of bad practices, with GPT-4o achieving the highest detection rate in our experiments. However, LLM selection is critical for detection performance, as models exhibit distinct strengths and limitations. \textsc{SCALM} can address these limitations through either multi-model specialization or single-model enhancement strategies.
\end{tcolorbox}

\subsection{RQ3: Comparison Experiments}
We conduct a comprehensive series of comparison experiments to address RQ3, which examines whether \textsc{SCALM} can identify bad practices that other tools cannot detect and how they compare with existing tools. These experiments are divided into two parts: \textbf{(1) Security-related Bad Practices Detection}, focusing on five critical SWC categories: SWC-101 (\emph{Integer Overflow and Underflow}), SWC-104 (\emph{Unchecked Call Return Value}), SWC-107 (\emph{Reentrancy}), SWC-112 (\emph{Delegatecall to Untrusted Callee}), and SWC-116 (\emph{Block Values as a Proxy for Time}), using 1,894 smart contracts from the SmartBugs dataset; and \textbf{(2) Quality-related Bad Practices Detection}, evaluating six categories of quality-related bad practices (Readability, Efficiency, Datatype, Function, Event, and Architecture) that affect code maintainability and design quality, using 1,200 smart contracts from the SolQuality dataset. In these experiments, \textsc{SCALM} is powered by GPT-4o, one of the most advanced LLMs.

Additionally, we collect a set of smart contract defect detection tools from reputable journals and conferences in software and security (e.g., CCS and ASE) as well as Mythril \cite{mythril}, recommended by the official Ethereum community.
For comparative analysis, we choose seven benchmark smart contract detection tools: four traditional tools (Mythril, Oyente \cite{Luu_2016_Making}, Confuzzius \cite{Torres_2021_ConFuzzius}, and Conkas \cite{VelosoConkas}) and three LLM-based tools (GPTLens \cite{hu2023large}, VulnHunt-GPT \cite{Boi_2024_VulnHunt-GPT}, and LLM-SmartAudit \cite{wei2025advanced}). Several factors are considered in the selection of the tools: \underline{(1)} The accessibility of the tool's source code. \underline{(2)} The tool's ability to detect the five categories of bad practices we select. \underline{(3)} The tool's support for source code written in Solidity. \underline{(4)} The tool's ability to report the exact location of potentially defective code for manual review. 
Note that traditional tools (Mythril, Oyente, Confuzzius, Conkas) focus exclusively on security-related bad practices and cannot evaluate quality-related aspects, while LLM-based tools support both dimensions.

\begin{table*}[h]
  \centering
  \small
    \caption{Comprehensive Performance Evaluation of Bad Practices Detection. Part I presents security-related bad practices detection results. Part II presents quality-related bad practices detection across six categories. LLM-SmartAudit employs the BA model for general auditing tasks.}
  \begin{minipage}[c]{0.48\textwidth}
    \begin{threeparttable}
    \centering
    \begin{tabular}{llccc}
    \toprule
    \multicolumn{5}{c}{\textbf{Part I: Security-related Bad Practices}} \\
    \midrule
    \textbf{SWC-ID} & \textbf{Tools} & \textbf{Acc} & \textbf{Rec} & \textbf{F1} \\
    \midrule
    \multirow{8}{*}{SWC-101} & Conkas & 49.27 & 67.20 & 59.10 \\
     & Mythril & 48.03 & 16.67 & 25.70 \\
     & Oyente & 62.33 & 69.35 & 66.03 \\
     & Confuzzius & 50.26 & 10.55 & 18.03 \\
     & GPTLens & 56.25 & 21.50 & 32.95 \\
     & VulnHunt-GPT & 55.50 & 29.50 & 39.86 \\
     & LLM-SmartAudit & 59.50 & 44.50 & 52.40 \\
     & \textbf{SCALM} & \textbf{95.50} & \textbf{94.50} & \textbf{95.45} \\
    \midrule
    \multirow{8}{*}{SWC-104} & Conkas & 59.88 & 20.12 & 33.00 \\
     & Mythril & 56.89 & 16.28 & 28.00 \\
     & Oyente & — & — & — \\
     & Confuzzius & 52.32 & 12.17 & 20.81 \\
     & GPTLens & 69.00 & 44.00 & 58.67 \\
     & VulnHunt-GPT & 53.25 & 8.50 & 15.38 \\
     & LLM-SmartAudit & 57.50 & 20.50 & 32.50 \\
     & \textbf{SCALM} & \textbf{98.25} & \textbf{99.50} & \textbf{98.27} \\
    \midrule
    \multirow{8}{*}{SWC-107} & Conkas & 71.79 & 94.51 & 77.50 \\
     & Mythril & 68.12 & 62.94 & 69.26 \\
     & Oyente & 59.12 & 14.04 & 24.49 \\
     & Confuzzius & 40.00 & 1.05 & 2.04 \\
     & GPTLens & 71.00 & 80.50 & 73.52 \\
     & VulnHunt-GPT & 52.00 & 62.50 & 56.56 \\
     & LLM-SmartAudit & 57.50 & 63.00 & 59.70 \\
     & \textbf{SCALM} & \textbf{95.00} & \textbf{94.50} & \textbf{94.97} \\
    \midrule
    \multirow{8}{*}{SWC-112} & Conkas & — & — & — \\
     & Mythril & 86.67 & 54.29 & 70.37 \\
     & Oyente & — & — & — \\
     & Confuzzius & 78.86 & 7.14 & 13.33 \\
     & GPTLens & 82.65 & 68.09 & 71.51 \\
     & VulnHunt-GPT & 81.97 & 43.62 & 60.74 \\
     & LLM-SmartAudit & 75.90 & 25.50 & 40.30 \\
     & \textbf{SCALM} & \textbf{98.30} & \textbf{95.74} & \textbf{97.30} \\
    \midrule
    \multirow{8}{*}{SWC-116} & Conkas & 89.35 & 86.99 & 88.50 \\
     & Mythril & 76.45 & 50.41 & 63.87 \\
     & Oyente & 48.20 & 3.16 & 6.03 \\
     & Confuzzius & — & — & — \\
     & GPTLens & 72.75 & 45.50 & 62.54 \\
     & VulnHunt-GPT & 60.25 & 25.50 & 39.08 \\
     & LLM-SmartAudit & 53.20 & 7.00 & 13.00 \\
     & \textbf{SCALM} & \textbf{93.00} & \textbf{88.50} & \textbf{92.67} \\
    \bottomrule
    \end{tabular}
    \end{threeparttable}
  \end{minipage}%
  \quad
  \begin{minipage}[c]{0.48\textwidth}
    \begin{threeparttable}
    \centering
    \begin{tabular}{llccc}
    \toprule
    \multicolumn{5}{c}{\textbf{Part II: Quality-related Bad Practices\tnote{*}}} \\
    \midrule
    \multicolumn{1}{l}{\textbf{Category}} & \multicolumn{4}{l}{\textbf{Category Descriptions}} \\
    \midrule
    \multicolumn{1}{l}{Readability} & \multicolumn{4}{p{0.7\linewidth}}{Poor naming, missing comments, messy structure} \\ 
    \multicolumn{1}{l}{Efficiency} & \multicolumn{4}{p{0.7\linewidth}}{Unnecessary storage, inefficient loops, redundant computations} \\
    \multicolumn{1}{l}{Datatype} & \multicolumn{4}{p{0.7\linewidth}}{Improper type selection, precision waste, redundant conversions} \\
    \multicolumn{1}{l}{Function} & \multicolumn{4}{p{0.7\linewidth}}{Non-single responsibility, excessive parameters, unclear returns} \\
    \multicolumn{1}{l}{Event} & \multicolumn{4}{p{0.7\linewidth}}{Missing events, unclear errors, improper exception handling} \\
    \multicolumn{1}{l}{Architecture} & \multicolumn{4}{p{0.7\linewidth}}{Code duplication, lack of modularity, confused inheritance} \\
    \midrule
    \textbf{Category} & \textbf{Tool} & \textbf{Acc} & \textbf{Rec} & \textbf{F1} \\
    \midrule
    \multirow{4}{*}{Readability} & GPTLens & — & — & —\\
     & VulnHunt-GPT & — & — & — \\
     & LLM-SmartAudit & 54.50 & 9.00 & 16.50 \\
     & \textbf{SCALM} & \textbf{91.50} & \textbf{83.00} & \textbf{90.71} \\
    \midrule
    \multirow{4}{*}{Efficiency} & GPTLens & 67.00 & 36.00 & 52.17 \\
     & VulnHunt-GPT & 69.50 & 43.00 & 58.50 \\
     & LLM-SmartAudit & 71.00 & 61.00 & 67.80 \\
     & \textbf{SCALM} & \textbf{89.00} & \textbf{85.00} & \textbf{88.54} \\
    \midrule
    \multirow{4}{*}{Datatype} & GPTLens & 52.00 & 4.00 & 7.69 \\
     & VulnHunt-GPT & 57.50 & 17.00 & 28.57 \\
     & LLM-SmartAudit & 60.50 & 22.00 & 35.80 \\
     & \textbf{SCALM} & \textbf{90.50} & \textbf{96.00} & \textbf{91.00} \\
    \midrule
    \multirow{4}{*}{Function} & GPTLens & 71.50 & 62.00 & 68.51 \\
     & VulnHunt-GPT & 56.50 & 13.00 & 23.01 \\
     & LLM-SmartAudit & 62.00 & 24.00 & 38.70 \\
     & \textbf{SCALM} & \textbf{88.00} & \textbf{97.00} & \textbf{88.99} \\
    \midrule
    \multirow{4}{*}{Event} & GPTLens & 45.50 & 8.00 & 12.80 \\
     & VulnHunt-GPT & 48.00 & 2.00 & 3.70 \\
     & LLM-SmartAudit & 58.00 & 57.00 & 57.60 \\
     & \textbf{SCALM} & \textbf{88.00} & \textbf{80.00} & \textbf{86.96} \\
    \midrule
    \multirow{4}{*}{Architecture} & GPTLens & — & — & — \\
     & VulnHunt-GPT & — & — & —\\
     & LLM-SmartAudit & 51.00 & 2.00 & 3.90 \\
     & \textbf{SCALM} & \textbf{91.50} & \textbf{94.00} & \textbf{91.71} \\
    \bottomrule
    \end{tabular}
    \begin{tablenotes}
    \footnotesize
    \item[*] LLM-based tools support both parts. Traditional tools focus only on security-related bad practices.
    \end{tablenotes}
    \end{threeparttable}
  \end{minipage}
  \label{table:comprehensive}
  \end{table*}

\textbf{Security-related Bad Practices Detection Results (Part I).} The experimental results in Table~\ref{table:comprehensive} demonstrate that \textsc{SCALM} achieves consistently high performance across all five SWC categories, with F1 scores ranging from 92.67\% to 98.27\%. Among traditional tools, Conkas and Oyente show competitive performance in specific categories: Conkas achieves 88.50\% F1 score for SWC-116, while Oyente reaches 66.03\% for SWC-101. Among LLM-based tools, GPTLens demonstrates relatively stable performance. Notably, for SWC-107 detection of Reentrancy, multiple tools, including Conkas, GPTLens, and Mythril, achieve F1 scores of 77.50\%, 73.52\%, and 69.26\%, respectively, though \textsc{SCALM}'s 94.97\% F1 score represents a notable improvement.

\textbf{Quality-related Bad Practices Detection Results (Part II).} The evaluation of quality-related bad practices reveals that \textsc{SCALM} maintains high performance across all six categories, with F1 scores exceeding 86.96\%. Among the baseline tools, performance varies considerably depending on the specific category. For Efficiency-related issues, LLM-SmartAudit achieves an F1 score of 67.80\%, while VulnHunt-GPT and GPTLens reach 58.50\% and 52.17\%, respectively. In Function-related detection, GPTLens demonstrates notable capability with an F1 score of 68.51\%. 
However, existing tools show limited effectiveness for more complex categories such as Readability, Datatype, and Architecture. The results suggest that while current LLM-based approaches can address certain quality-related issues, they face challenges in detecting more nuanced bad practices that require deeper code understanding and architectural analysis.

The performance of GPTLens, VulnHunt-GPT, and LLM-SmartAudit shows notable differences compared to SCALM, which several methodological factors can explain. First, their knowledge bases and prompts tend to emphasize security aspects like "exploitable vulnerabilities" and "potential attack vectors" rather than code quality metrics, including naming conventions, modularity, or maintainability. Second, each tool has distinct architectural characteristics: GPTLens implements a conservative criticism mechanism with strict evaluation policies, which may filter certain quality-related issues; VulnHunt-GPT's RAG approach operates with a knowledge base covering only seven vulnerability types; LLM-SmartAudit utilizes <INFO> tags from the LLM for decision-making. In comparison, SCALM employs a multi-layer reasoning validation mechanism (syntax, design patterns, architecture) and abstract reasoning guided by Step-Back prompting, enabling it to address security vulnerabilities and code quality issues.

\begin{tcolorbox}[boxrule=1pt,boxsep=1pt,left=2pt,right=2pt,top=1pt,bottom=1pt]
\textbf{Answer to RQ3.}
\textsc{SCALM} demonstrates improved detection performance compared to existing tools across security-related and quality-related bad practices. These results demonstrate that \textsc{SCALM}'s multi-layer reasoning mechanism enables it to detect low-level syntax issues and high-level design problems, providing more comprehensive analysis than existing tools.
\end{tcolorbox}

\subsection{RQ4: Ablation Experiments}
We conduct ablation experiments to address RQ4, investigating whether \textsc{SCALM} can achieve the same performance without including the RAG component. In these experiments, \textsc{SCALM} is also powered by GPT-4o with the same evaluation setup as RQ3. We compare three configurations: \textbf{SCALM} (full framework with all components), SCALM\textsuperscript{-R} (excluding the RAG component), and SCALM\textsuperscript{-R-M} (excluding both RAG and multi-layer reasoning). The results are summarized in Table~\ref{table:swc_rag}.

\begin{table}[htbp]
  \centering
  \small
    \caption{Ablation Experiments. 'SCALM\textsuperscript{-R-M}' indicates excluding all components, 'SCALM\textsuperscript{-R}' indicates excluding the RAG component, while SCALM includes all components.}
  \begin{threeparttable}
  \begin{minipage}[t]{0.495\linewidth}
    \centering
    \begin{tabular}{lcccc}
    \toprule
    \textbf{Bad Practice}  & \textbf{Tools} & \textbf{Acc} & \textbf{Rec}  & \textbf{F1} \\
    \midrule
    \multirow{3}{*}{SWC-101}  & SCALM\textsuperscript{-R-M} & 69.25 & 87.00 & 73.89 \\[2.31pt]
     & SCALM\textsuperscript{-R} & 79.25 & 81.50 & 79.71 \\[2.31pt]
     & \textbf{SCALM} & \textbf{95.50} & \textbf{94.50}  & \textbf{95.45} \\[2.31pt]
    \midrule
    \multirow{3}{*}{SWC-104}  & SCALM\textsuperscript{-R-M} & 54.00 & 17.50 & 27.56 \\[2.31pt]
     & SCALM\textsuperscript{-R} & 77.00 & 87.50  & 79.19 \\[2.31pt]
     & \textbf{SCALM} & \textbf{98.25} & \textbf{99.50}  & \textbf{98.27} \\[2.31pt]
    \midrule
    \multirow{3}{*}{SWC-107}  & SCALM\textsuperscript{-R-M} & 58.50 & 71.00 & 63.11 \\[2.31pt]
     & SCALM\textsuperscript{-R} & 73.50 & 81.00  & 75.35 \\[2.31pt]
     & \textbf{SCALM} & \textbf{95.00} & \textbf{94.50}  & \textbf{94.97} \\[2.31pt]
    \midrule
    \multirow{3}{*}{SWC-112}  & SCALM\textsuperscript{-R-M} & 76.53 & 41.49 & 53.06 \\[2.31pt]
     & SCALM\textsuperscript{-R} & 84.69 & 68.09  & 73.99 \\[2.31pt]
     & \textbf{SCALM} & \textbf{98.30} & \textbf{95.74}  & \textbf{97.30} \\[2.31pt]
    \midrule
    \multirow{3}{*}{SWC-116}  & SCALM\textsuperscript{-R-M} & 52.50 & 11.00 & 18.80 \\[2.31pt]
     & SCALM\textsuperscript{-R} & 82.75 & 85.00 & 83.13 \\[2.31pt]
     & \textbf{SCALM} & \textbf{93.00} & \textbf{88.50}  & \textbf{92.67} \\[2.31pt]
    \bottomrule
    \end{tabular}
  \end{minipage}%
  \hspace{0.5em}
  \begin{minipage}[t]{0.495\linewidth}
    \centering
    \begin{tabular}{lcccc}
    \toprule
    \textbf{Bad Practice}  & \textbf{Tools} & \textbf{Acc} & \textbf{Rec}  & \textbf{F1} \\
    \midrule
    \multirow{3}{*}{Readability} & SCALM\textsuperscript{-R-M} & 51.50 & 3.00 & 5.83 \\
    & SCALM\textsuperscript{-R} & 71.50 & 43.00 & 60.14 \\
    & \textbf{SCALM} & \textbf{91.50} & \textbf{83.00} & \textbf{90.71} \\
    \midrule
    \multirow{3}{*}{Efficiency} & SCALM\textsuperscript{-R-M} & 86.00 & 82.00 & 85.42 \\
    & SCALM\textsuperscript{-R} & 87.50 & 84.00 & 87.05 \\
    & \textbf{SCALM} & \textbf{89.00} & \textbf{85.00} & \textbf{88.54} \\
    \midrule
    \multirow{3}{*}{Datatype} & SCALM\textsuperscript{-R-M} & 56.50 & 18.00 & 29.27 \\
    & SCALM\textsuperscript{-R} & 73.50 & 57.00 & 68.26 \\
    & \textbf{SCALM} & \textbf{90.50} & \textbf{96.00} & \textbf{91.00} \\
    \midrule
    \multirow{3}{*}{Function} & SCALM\textsuperscript{-R-M} & 62.00 & 75.00 & 66.37 \\
    & SCALM\textsuperscript{-R} & 75.00 & 86.00 & 77.48 \\
    & \textbf{SCALM} & \textbf{88.00} & \textbf{97.00} & \textbf{88.99} \\
    \midrule
    \multirow{3}{*}{Event} & SCALM\textsuperscript{-R-M} & 70.50 & 57.00 & 65.90 \\
    & SCALM\textsuperscript{-R} & 79.00 & 68.00 & 76.40 \\
    & \textbf{SCALM} & \textbf{88.00} & \textbf{80.00} & \textbf{86.96} \\
    \midrule
    \multirow{3}{*}{Architecture} & SCALM\textsuperscript{-R-M} & 56.00 & 26.00 & 37.14 \\
    & SCALM\textsuperscript{-R} & 74.00 & 60.00 & 69.77 \\
    & \textbf{SCALM} & \textbf{91.50} & \textbf{94.00} & \textbf{91.71} \\
    \bottomrule
    \end{tabular}
  \end{minipage}
  \end{threeparttable}

  \label{table:swc_rag}
\end{table}

For both security and quality-related bad practices detection, RAG augmentation demonstrates significant performance improvements across different pattern categories. For security-related bad practices, the average F1 score improvement is 17.5\% compared to SCALM\textsuperscript{-R}, with SWC-104 and SWC-112 showing the largest absolute gains, achieving 98.27\% and 97.30\% F1 scores, respectively. SWC-116 exhibits a more modest improvement at 92.67\% F1, though still representing a 9.54\% increase. For quality-related bad practices, the improvements are even more pronounced: Readability detection improves from 60.14\% to 90.71\% F1, Datatype detection jumps from 68.26\% to 91.00\%, and Architecture-level analysis shows the most dramatic gain from 69.77\% to 91.71\%. Notably, patterns requiring complex semantic validation (SWC-104, SWC-112, Datatype, Architecture) show more substantial improvements, while self-contained patterns (Efficiency) exhibit relatively modest gains, highlighting RAG's critical role in high-level reasoning tasks.

The multi-layer reasoning component contributes to detection performance. Comparing SCALM\textsuperscript{-R} with SCALM\textsuperscript{-R-M}, we observe an average F1 improvement of 30.99\% across security patterns and 24.86\% across quality patterns. The impact varies across different pattern types: SWC-104 shows a 51.63\% F1 improvement, SWC-112 improves by 20.93\%, and Readability detection increases by 54.31\%. These results suggest that multi-layer reasoning benefits patterns requiring semantic analysis at multiple abstraction levels. Patterns with more explicit detection criteria (e.g., Efficiency) show more minor improvements of 1.63\%, indicating that the effectiveness of multi-layer reasoning depends on the complexity and nature of the detection task.


\begin{tcolorbox}[boxrule=1pt,boxsep=1pt,left=2pt,right=2pt,top=1pt,bottom=1pt]
\textbf{Answer to RQ4.}
Ablation experiments demonstrate that both RAG and multi-layer reasoning contribute to the performance of \textsc{SCALM}. RAG enhances detection by providing contextual evidence and reducing false positives through cross-referencing. Multi-layer reasoning aids in decomposing the detection task into subproblems, benefiting patterns requiring semantic understanding. Combining these two components improves the performance of SCALM, supporting the design of the complete framework architecture.
\end{tcolorbox}

\section{Discussion}
The experimental results confirm \textsc{SCALM}'s effectiveness in detecting bad practices in smart contracts, outperforming existing tools in accuracy, recall, and F1 scores.
The choice of LLM significantly impacts \textsc{SCALM}'s performance, with different models yielding varying results. This finding emphasizes the need for careful model selection and ongoing improvements to maintain high accuracy. Ablation experiments showed significant performance drops when RAG and multi-layer reasoning components were excluded. This highlights its importance in providing contextual information that enhances detection accuracy.

Beyond performance improvements, the multi-layer reasoning verification mechanism represents a paradigm shift from traditional pattern-matching approaches to semantic understanding in smart contract analysis. Our framework's ability to abstract from syntax-level patterns to architectural principles demonstrates that LLMs can bridge the gap between low-level code vulnerabilities and high-level security design principles. This hierarchical abstraction capability suggests potential applications beyond smart contract auditing, including general software architecture analysis and security-by-design validation.

Despite its strengths, \textsc{SCALM} has limitations in understanding blockchain mechanisms and complex interactions \cite{zhang2025attacks,zhang2025security}. Specifically, \textsc{SCALM} might not fully grasp the state-dependent nature of certain vulnerabilities, which require a deep contextual understanding of how the contract evolves over time. 
Future work will focus on improving \textsc{SCALM} by incorporating advanced LLMs and fine-tuning with domain-specific data. Additionally, we plan to enhance the RAG generation strategy to provide more prosperous and accurate contextual information. These improvements will improve the performance of \textsc{SCALM} for smart contract auditing.

\section{Related Work}
LLMs have been widely applied and validated for their ability to identify and fix vulnerabilities \cite{Napoli_2023_Evaluating,10.1145/3643674}. In the field of smart contract security, various methods based on LLMs have been proposed, and certain effects have been achieved.
Firstly, Boi et al. \cite{Boi_2024_VulnHunt-GPT} proposed VulnHunt-GPT, a method that uses GPT-3 to identify common vulnerabilities in OWASP-standard smart contracts. Building on this, Boi et al. \cite{10.1145/3687251.3687253} further demonstrated how LLMs trained on diverse vulnerability datasets can detect multiple security flaws simultaneously, outperforming traditional static analysis tools. Similarly, Xia et al. \cite{Xia_2024_AuditGPT} introduced AuditGPT, which utilizes LLMs to verify ERC rules through decomposed audit tasks. In terms of fuzz testing, Shou et al. \cite{Shou_2024_LLM4Fuzz} proposed LLM4Fuzz to guide fuzz testing priorities, while Zhao et al. \cite{zhao2025detecting} enhanced functional bug detection by combining LLM-driven vulnerability pinpointing with bug-oriented fuzzing. Ding et al. \cite{DING2025126479} developed SmartGuard, leveraging LLMs for semantic code retrieval and Chain-of-Thought reasoning to achieve 95\% F1-scores on benchmarks. Wei et al. \cite{wei2024llm} advanced multi-agent collaborative auditing with LLM-SmartAudit, detecting complex logic vulnerabilities overlooked by traditional tools.

However, the direct use of pre-trained LLMs is no longer sufficient on some occasions, so many studies have chosen to fine-tune LLMs to meet specific needs. For example, Luo et al. \cite{10664408} proposed FELLMVP, an ensemble framework that fine-tunes eight LLMs for specific vulnerabilities and integrates their predictions, achieving 98.8\% accuracy. Mothukuri et al. \cite{10664261} introduced LLMSmartSec, which fine-tunes GPT-4 to analyze contracts from developer, auditor, and ethical hacker perspectives, then distills knowledge into an LLMGraphAgent via annotated control flow graphs. Liu et al. \cite{Liu_2024_PropertyGPT} designed PropertyGPT for formal verification by generating contract properties through retrieval-augmented LLMs. Storhaug et al. \cite{Storhaug_2023_Efficient} reduced vulnerable code generation via vulnerability-bound decoding, while Yang et al. \cite{Yang_2024_Automated} fine-tuned Llama-2-13B for DeFi-specific vulnerability detection.
In conclusion, while LLMs show great potential for improving smart contract security, the full realization of their potential requires continuous fine-tuning and adaptation of these models in specific contexts.

\section{Conclusion}
This paper presents the first systematic study of over 47 bad practices in smart contracts. Building on this extensive analysis, we introduced \textsc{SCALM}, a framework based on LLMs for detecting these bad practices in smart contracts. It extracts bad practice patterns and builds an extensible knowledge base through context-aware function-level slicing, combines RAG and Step-Back prompting for multi-layer reasoning and validation, and ultimately generates structured audit reports.
Experiments on the SmartBugs and solquality datasets demonstrate that \textsc{SCALM} achieves superior performance compared to existing tools, with F1 scores exceeding 92\% across five SWC categories and 86\% across six quality dimensions. Ablation studies confirm the essential contributions of both RAG and multi-layer reasoning components, with RAG providing 17.5\% average F1 improvement for security patterns and multi-layer reasoning contributing 30.99\% additional gains. These findings demonstrate \textsc{SCALM}'s potential to enhance smart contract security, offering developers a robust framework to identify and fix bad practices.

\section*{Acknowledgments}
This work is sponsored by the National Natural Science Foundation of China (No.62402146 \& 62362021). 

\bibliography{tosem}

\end{document}